\documentclass[aps,prd,twocolumn,nofootinbib,floatfix,notitlepage,superscriptaddress]{revtex4-2}
\usepackage{amsmath,amsfonts,amssymb}
\usepackage{mathrsfs}
\usepackage{multirow}
\usepackage{graphicx}
\usepackage[english]{babel}
\usepackage{url}
\usepackage{color}
\usepackage[utf8]{inputenc}
\usepackage[colorlinks=true,citecolor=blue,urlcolor=blue]{hyperref}
\usepackage{booktabs}
\usepackage[T1]{fontenc}

\begin{document}

\title{Geometric obstruction to resolving the Hubble tension:\\ orthogonality of scale and shape in distance measurements}

\author{Zhihuan Zhou}
\email{zhihuanzhou1@163.com}
\author{ShengYue Wang}%
\author{Zhuang Miao}%
\author{Chaoqian Ai}%
\affiliation{
	School of Engineering \\
	Xi'an International University \\
	Xi'an 710077, People's Republic of China}%

\author{Hongchao Zhang}
\email{zhanghongchao852@live.com}
\affiliation{
	Department of Physics \\
	Liaoning Normal University \\
	Dalian, 116029, People's Republic of China}%

\date{\today}

\begin{abstract}
	We identify a geometric obstruction to resolving the Hubble tension by combining early-time sound-horizon reduction with late-time smooth dark energy.  Within $\Lambda$CDM, the BAO--SN matter-density gap $\Delta\Omega_m = 0.037$ is exactly invariant under the sound-horizon rescaling $\alpha \equiv r_s^{\rm mod}/r_s^{\Lambda{\rm CDM}}$, and late-time $w(z)$ deformations cannot eliminate this gap either: reconciling the two datasets requires \emph{opposite} deformations---phantom ($w < -1$) for BAO, quintessence ($w > -1$) for SN at $z < 0.5$---an anti-alignment quantified by $\cos\theta = -0.97$ in $w(z)$ space.  A full MCMC analysis of DESI DR2 BAO, Planck plik\_lite, and Pantheon+ bears this out: the optimal $\alpha^* = 0.992$ ($0.8\%$ $r_s$ reduction) brings the joint fit to $H_0 = 70.3 \pm 0.3\;\mathrm{km\,s^{-1}\,Mpc^{-1}}$, still $3.2\sigma$ below SH0ES, with the inter-dataset tension reduced but not removed.  The obstruction reflects not a shortage of model freedom but an irreducible disagreement between probes.  The deformation space $\{\alpha, \beta_{\rm damp}, w(z)\}$ already spans $93\%$ of the $\Omega_m$ response direction; nonetheless BAO and SN constrain $\Omega_m$ through independent channels and disagree, while the residual $H_0$ deficit, anchored by the local distance ladder, resides in the absolute distance scale that $w(z)$ reshapes but cannot rescale.
\end{abstract}
\maketitle

\section{Introduction}\label{sec:intro}

The Hubble tension is among the most persistent challenges in modern cosmology \cite{Verde:2023lmm,Verde:2019ivm,DiValentino:2021izs,Perivolaropoulos:2021jda,CosmoVerse:2025txj,Kamionkowski:2022pkx}.  It refers to a $>5\sigma$ discrepancy between the Hubble constant inferred from the \textit{Planck} cosmic microwave background (CMB), $H_0 = 67.36 \pm 0.54\;\mathrm{km\,s^{-1}\,Mpc^{-1}}$ \cite{Aghanim:2018eyx}, and the local SH0ES distance ladder, $H_0 = 73.17 \pm 0.86\;\mathrm{km\,s^{-1}\,Mpc^{-1}}$ \cite{Riess:2024vfa}.  The discrepancy is reinforced by independent local measurements, including strong-lensing time delays from H0LiCOW \cite{Wong:2019kwg} and sound-horizon-free BAO analyses \cite{Zaborowski:2024car,Pantos:2026cxv}, and persists across evolving datasets.  Its resolution has motivated a wide range of proposed modifications to $\Lambda$CDM.

Early-time approaches modify pre-recombination physics to reduce the sound horizon $r_s$, thereby increasing $H_0$ at fixed CMB angular scales \cite{Knox:2019rjx,Schoneberg:2021qvd}.  Concrete mechanisms include early dark energy (EDE) \cite{Poulin:2023lkg,Hill:2020osr}, modified recombination \cite{Lee:2022gzh,Mirpoorian:2024fka,Jedamzik:2025cax}, and extra radiation species \cite{Vagnozzi:2021gjh}.  While some of these models can raise $H_0$ closer to the local value, they face several well-known difficulties.  Jedamzik \textit{et al.}~\cite{Jedamzik:2020zmd} showed that reducing $r_s$ alone cannot fully resolve the tension, because the required shift introduces compensating effects in other parameters.  EDE tends to worsen the $S_8$ tension---which manifests not only in weak lensing but also in growth rate measurements from redshift-space distortions \cite{Nunes:2021ipq}---as shown by several analyses \cite{Vagnozzi:2019ezj,Pedrotti:2024kpn}.  Achieving full concordance requires $r_s$ reductions of $\sim$5--$8\%$ that produce detectable distortions in the CMB damping tail \cite{Lee:2025yah,Vagnozzi:2023nrq,Pedrotti:2026dwj}.  Early-time solutions also generically introduce a $\omega_b$ tension \cite{Giovanetti:2026aku}.  Moreover, even combining early-time new physics with late-time flexibility leaves a residual ``cosmic calibration tension'' that extends beyond $H_0$ alone \cite{Poulin:2024ken}.

Late-time approaches instead modify the dark energy equation of state $w(z)$ or the expansion rate $H(z)$ at $z \lesssim 3$ \cite{Benevento:2020fev,Efstathiou:2021ocp,Yang:2018qmz,Heisenberg:2022gqk,Escamilla:2023oce,Zhou:2021xov,Scherer:2025esj,Dai:2020rfo,DeSimone:2024lvy,Navone:2025gxr,Valletta:2025bgu,Silva:2025hxw,Adi:2025hyj,Hussain:2025nqy}.  However, formal no-go results \cite{Cai:2021weh,Huang:2024erq} and SN distance measurements tightly constrain luminosity distances, leaving little room for the large $H(z)$ modifications needed to bridge the $H_0$ gap \cite{Benevento:2020fev,Gomez-Valent:2023uof,Pedrotti:2025ccw,Colgain:2021beg}, and late-time $H(z)$ deformations that raise $H_0$ tend to worsen the growth tension \cite{Alestas:2021xes}.  DESI BAO data have been interpreted as favoring evolving dark energy \cite{DESI:2025zgx,DESI:2025fii,Jia:2025poj,Wang:2025xvi}, but several analyses have questioned whether this evidence is robust \cite{Wang:2025bkk,Huang:2025som,Pang:2025lvh}, and the preferred $w_0w_a$CDM parameters do not resolve the $H_0$ discrepancy \cite{Zhang:2025lam} but instead introduce $\Omega_m(z)$ inconsistencies between probes \cite{OColgain:2024xqj,Afroz:2025iwo}.  Neither class of modification alone has proven sufficient, and a natural question is whether combining both freedoms provides the flexibility needed for concordance.

A key insight emerging from multiple independent analyses is that the Hubble tension contains an irreducible $\Omega_m$ component, persistent across modification channels.  In our previous work \cite{Zhou:2024dbt}, we showed that the $w(z)$-space gradients of BAO and SN $\chi^2$ are anti-aligned, preventing any smooth dark energy modification from simultaneously satisfying both probes.  Shlivko and Poulin~\cite{Shlivko:2026jxa} traced the observational preference for phantom crossing directly to the $\Omega_m$ ordering between BAO, CMB, and SN datasets, and the $\Omega_m$ prior bias between DESI DR2 BAO and Pantheon+ has been explicitly quantified \cite{Lee:2025kbn}.

Crucially, early-time and geometric solutions leave this $\Omega_m$ gap intact.  Lee \textit{et al.}~\cite{Lee:2025yah} found that even optimizing the full primordial power spectrum $P(k)$ achieves $H_0 = 73$ only at $\Omega_m = 0.247$, in severe tension with BAO.  Modified recombination likewise leaves the $\Omega_m$ conflict intact \cite{Mirpoorian:2024fka,Mirpoorian:2025rfp}, while Bansal and Huterer \cite{Bansal:2026late} confirmed via MCMC that smooth $H(z)$ modifications yield negligible improvement---only SN calibration transitions produce significant relief.  Pedrotti \textit{et al.}~\cite{Pedrotti:2024kpn} showed that the structural link $\delta\omega_c/\omega_c \approx 2.38\,\delta h/h$ constrains Planck's $\Omega_m$--$H_0$ degeneracy, making it impossible to change one without the other.

Further support comes from redshift trends and information-theoretic arguments.  Binned analyses of SN Ia samples reveal a decreasing trend of $H_0$ with redshift \cite{Dainotti:2021pqg,Dainotti:2022bzg,Dainotti:2025qxz}, consistent with a redshift-dependent $\Omega_m$ discrepancy, and similar trends appear in $\Lambda$CDM parameter tracking across redshift bins \cite{Krishnan:2021dyb,Colgain:2024clf,Colgain:2024ksa}.  On the information-theoretic side, Lee \cite{Lee:2026geo} showed that extending $\Lambda$CDM to wCDM reduces the leading Planck Fisher eigenvalue by a factor of $37.5$, indicating severe compression of cosmological information.  CMB-based constraints on allowed $H(z)$ deviations from $\Lambda$CDM further limit the available parameter space \cite{Das:2023rvg}, and the tension has been reformulated in terms of the full expansion function $E(z)$ \cite{Lee:2026bxq}.  Attempts to reconcile the SN--BAO discrepancy through inhomogeneous-universe models \cite{Futamase:2026zos} or by tracing the origin of the BAO tension \cite{Pantos:2026rpe} further underscore the depth of the conflict.  These analyses, together with the studies above, highlight different manifestations of an underlying inconsistency between distance probes---in both $\Omega_m$ and $H_0$---that persists across methods and datasets.

In this paper, we identify the structural reason for this persistence.  Our central observation is that $r_s$ rescaling and $w(z)$ deformations affect \emph{structurally distinct} aspects of distance measurements.  The parameter $\alpha \equiv r_s^{\rm mod}/r_s^{\Lambda\rm CDM}$ operates on the \emph{absolute distance scale}: it sets the ruler against which BAO measures distances.  But $\Omega_m$ is determined by \emph{relative distance ratios} $D(z_2)/D(z_1)$, in which the ruler cancels.  This means the BAO--SN $\Omega_m$ gap is exactly $\alpha$-invariant---no early-time mechanism operating through $r_s$ reduction can touch it.  Meanwhile, $w(z)$ can modify relative distances, but BAO and SN demand modifications in opposite directions.  Geometrically, the three datasets sit at three separated points in the $(\Omega_m, H_0)$ plane (shown later in Fig.~\ref{fig:tension_surface}).  Lowering $\alpha$ slides BAO straight up in $H_0$ and leaves SN fixed, so it cannot change the BAO--SN $\Omega_m$ gap; Planck, by contrast, moves along a diagonal, gaining a higher $H_0$ only at the cost of a lower $\Omega_m$.  Late-time $w(z)$ can shift points horizontally in $\Omega_m$, but BAO and SN must move in \emph{opposite} directions.  No combination of the two brings all three points together---that is the obstruction.  We verify these geometric properties with full MCMC chains using DESI DR2 BAO, Planck plik\_lite, and Pantheon+ supernovae, fitting all six $\Lambda$CDM parameters simultaneously.

The paper is organized as follows.  Section~\ref{sec:method} describes the analysis framework and datasets.  Section~\ref{sec:theorem} presents the geometric obstruction: scale--shape decoupling and $w(z)$ anti-alignment.  Section~\ref{sec:results} provides MCMC verification.  Section~\ref{sec:wz} demonstrates that $w(z)$ cannot close the $\Omega_m$ gap.  Section~\ref{sec:freedom} analyzes the effective freedom of the deformation space.  Section~\ref{sec:robust} tests robustness.  We discuss converging evidence and implications in Sec.~\ref{sec:discussion} and conclude in Sec.~\ref{sec:conclusions}.

\section{Method}\label{sec:method}

\subsection{Analysis framework}\label{sec:framework}

We employ the Fisher-bias perturbation analysis (FPA) framework of Ref.~\cite{Zhou:2024dbt} for analytical gradient computations, validated by full MCMC chains for all quantitative claims (Sec.~\ref{sec:results}).

The FPA framework computes the response of best-fit parameters to $w(z)$ deformations through
\begin{equation}
	\Delta \Omega_{\rm BF}^i = -(F^{-1})_{ij} \frac{\partial \bm{X}}{\partial \Omega^j} \cdot \bm{\Sigma}^{-1} \cdot \frac{\delta \bm{X}}{\delta w(z)} \cdot \bm{c}, \label{eq:response}
\end{equation}
where $F_{ij}$ is the Fisher matrix and $\bm{X}$ is the observable vector.  We use FPA for gradient directions and response-vector analysis, evaluated at the MCMC best-fit point.  The directional conclusions (anti-alignment sign, response overlap) depend on the dataset geometry, not on the number of free parameters.

Inter-dataset consistency is quantified by
\begin{equation}
	T(\alpha) \equiv \chi^2_{\rm joint}(\alpha) - \sum_d \chi^2_{d,\min}(\alpha), \label{eq:tension}
\end{equation}
where $\chi^2_{d,\min}$ is each dataset's minimum $\chi^2$ with all parameters free, and $\chi^2_{\rm joint}$ is the minimum when all datasets are fit simultaneously.  In words, $T$ measures how much worse the joint fit is than fitting each dataset on its own; $T = 0$ means perfect agreement.  By construction $T \geq 0$, with $T = 0$ if and only if all datasets prefer identical parameter values.  For $N_d = 3$ datasets sharing $k$ constrained parameters, $T$ approximately follows a $\chi^2$ distribution with $(N_d - 1) \times k$ degrees of freedom under the null hypothesis of dataset agreement (assuming Gaussian likelihoods near the best fit).  With $k = 2$ effective parameters ($\Omega_m$, $H_0$), the approximate null distribution is $\chi^2(4)$.

We also define pairwise tensions $T_{AB} \equiv \min_\theta[\chi^2_A(\theta) + \chi^2_B(\theta)] - \chi^2_{A,\min} - \chi^2_{B,\min}$, which follow $\chi^2(k)$ under the null.  Throughout, we convert tension values to significance levels via the exact $\chi^2 \to p\text{-value} \to$ Gaussian-equivalent mapping.

\subsection{Model-independent deformation basis}\label{sec:basis}

\paragraph{Early-time: $\alpha$ and $\beta_{\rm damp}$.}
The CMB spectral effects of $r_s$-reducing mechanisms are captured at leading order by two phenomenological parameters:
\begin{equation}
	C'_\ell = C_{\ell \cdot \alpha} \cdot \exp(-\beta_{\rm damp}\, \ell^2), \label{eq:alpha_beta}
\end{equation}
where $\alpha \equiv r_s^{\rm mod}/r_s^{\Lambda{\rm CDM}}$ shifts peak positions ($\ell_n \propto 1/r_s$) and $\beta_{\rm damp}$ parametrizes Silk damping modifications from recombination physics \cite{Mirpoorian:2024fka}.  For BAO, $\alpha$ rescales the standard ruler: $D(z_i)/r_d \to D(z_i)/(\alpha \cdot r_d)$.  SN luminosity distances are $r_s$-independent and therefore exactly $\alpha$-independent.

\paragraph{Late-time: $w(z)$.}
Expansion-history modifications are parametrized through 20 Gaussian basis functions $w(z) = -1 + \sum_{j=1}^{20} c_j \phi_j(z)$ with nodes equally spaced over $z \in [0, 3]$.

\paragraph{Datasets.}
The combined analysis uses 2206 data points: 13 DESI DR2 BAO distance measurements \cite{DESI:2025zgx}, 613 Planck plik\_lite bins (TT+TE+EE, $30 \leq \ell \leq 2508$) \cite{Planck:2019nip}, and 1580 Pantheon+ SN distance moduli \cite{Scolnic:2021amr,Brout:2022vxf} with the SH0ES Cepheid calibration of the absolute magnitude (this calibrated Pantheon+ sample is denoted simply \emph{SN} hereafter).  We analytically marginalize $M_B$ with a Gaussian SH0ES prior $M_B = -19.253 \pm 0.027$; this calibration anchors the absolute distance scale, so SN constrains \emph{both} $\Omega_m = 0.334$ and $H_0 = 73.2 \pm 1.0\;\mathrm{km\,s^{-1}\,Mpc^{-1}}$ (Table~\ref{tab:bestfits}).  The SN central $H_0$ thus inherits the SH0ES calibration; its $\pm 1.0$ posterior is the width of the SN chain and is broader than the SH0ES prior itself ($73.17 \pm 0.86$ \cite{Riess:2024vfa}), against whose tighter error we quote the residual $H_0$ significances below.  Lensed $C_\ell$ are computed with CLASS \cite{Blas:2011rf}; at the fiducial, $\chi^2/{\rm dof} = 549/613$.  To test robustness, we also use DES-SN5YR \cite{DES:2024jxu} (1820 SNe).

\section{The geometric obstruction}\label{sec:theorem}

The central result of this paper is that $r_s$ rescaling and the determination of $\Omega_m$ are geometrically decoupled within the space of smooth FLRW expansion histories.  The argument relies on the following assumptions: (i) a smooth Friedmann--Lema\^itre--Robertson--Walker background, (ii) standard BAO ruler interpretation ($D_V(z)/r_d$), (iii) SN luminosity distances containing no $r_s$ dependence, (iv) sound horizon modifications entering only through $\alpha$, and (v) no perturbation-level effects that modify the distance--$\Omega_m$ mapping.

The obstruction operates at two levels.  \emph{First}, $\alpha$ cannot close the BAO--SN $\Omega_m$ gap because it acts on the absolute distance scale while $\Omega_m$ is fixed by relative ratios in which $\alpha$ cancels.  \emph{Second}, even the remaining $w(z)$ freedom cannot close the gap because BAO and SN require opposite deformations (Sec.~\ref{sec:scale_shape}).  The two levels are independent: the first is a kinematic identity of distance observables, while the second is an empirical property of the current data.  Sec.~\ref{sec:results} provides MCMC verification of both statements.

\subsection{Scale versus shape in BAO distances}\label{sec:scale_shape}

The BAO distance observable $D_V(z_i)/(\alpha \cdot r_d)$ factorizes into two structurally independent components.  The key property is that $\alpha$ cancels exactly in any distance ratio, regardless of the expansion history.  Writing $D_V(z) = (c/H_0)\, g(\Omega_m, w(z), z)$ for a general smooth dark energy model, the ratio $D_V(z_2)/D_V(z_1) = g(\Omega_m, w(z), z_2)/g(\Omega_m, w(z), z_1)$ contains no dependence on $H_0$, $\alpha$, or $r_d$.  Sound horizon rescaling therefore cannot alter the shape information from which $\Omega_m$ is inferred; it enters only through the overall normalization $D_V(z)/(\alpha \cdot r_d) \propto 1/(H_0 \cdot \alpha \cdot r_d)$, determining $H_0(\alpha)$ at fixed shape.

Within $\Lambda$CDM ($w = -1$), the shape function simplifies to
\begin{equation}
	g(\Omega_m, z) = \left[\frac{z}{E(z)} \left(\int_0^z \frac{dz'}{E(z')}\right)^2\right]^{1/3},
\end{equation}
with $E(z) = [\Omega_m(1+z)^3 + 1-\Omega_m]^{1/2}$.  In this $\Lambda$CDM limit the general shape function specializes to depend on $\Omega_m$ alone, so that $\Omega_m$ becomes the sole parameter governing relative distances and---given measurements at multiple redshifts---BAO determines it from shape alone, while $\alpha$ continues to enter only through the overall scale $D_V(z)/(\alpha \cdot r_d)$.  The SN distance modulus $\mu(z_i) = 5\log_{10}[d_L(z_i)/10\,\text{pc}]$ depends on $d_L(z) = (c/H_0)(1+z)\int_0^z dz'/E(z')$, which contains no $r_s$.  The SN $\chi^2$ is therefore completely $\alpha$-independent---both the $\Omega_m$ and $H_0$ inferred from SN are unaffected by $r_s$ rescaling, and this $\alpha$-independence holds regardless of the $M_B$ prior.  Therefore:
\begin{equation}
	\Delta\Omega_m(\text{BAO}, \text{SN})\big|_{w=-1} \approx 0.037 \quad \forall\;\alpha. \label{eq:invariant}
\end{equation}
Within $\Lambda$CDM, no early-time model operating through $r_s$ reduction---early dark energy, extra radiation, modified recombination, or any other mechanism---can close this gap.  When the late-time expansion is simultaneously modified by $w(z) \neq -1$, the shape function $g$ depends on both $\Omega_m$ and $w(z)$, breaking the strict $\Omega_m$-only determination.  However, this additional freedom does not help: BAO and SN require opposite $w(z)$ modifications.  At the MCMC joint best-fit ($\alpha = 0.992$, $H_0 = 70.3$, $\Omega_m = 0.290$), the 20-dimensional $w(z)$-space gradients satisfy
\begin{equation}
	\cos(g_{\rm BAO}, g_{\rm SN}) = -0.97, \label{eq:antialign}
\end{equation}
meaning that $w(z)$ deformations reducing BAO $\chi^2$ tend to increase SN $\chi^2$ by a comparable amount along the anti-aligned component.  This anti-alignment persists at every $\alpha$ tested, because it reflects the underlying data conflict rather than the evaluation point.

\subsection{Factorization and its observational consequences}\label{sec:factorize}

The two properties above---$\alpha$-invariance of $\Delta\Omega_m$ and $w(z)$ anti-alignment---imply that the joint optimization over $(\alpha, w(z))$ decomposes into two independent subproblems.  Varying $\alpha$ optimizes along $H_0$: it adjusts the absolute distance scale, balancing the Planck--SN and Planck--BAO tensions without affecting $\Delta\Omega_m$.  The equation-of-state freedom $w(z)$, at any fixed $\alpha$, instead optimizes along $\Omega_m$, attempting to reconcile the residual gap by reshaping the distance--redshift relation; and because $\Delta\Omega_m$ is exactly $\alpha$-invariant [Eq.~\eqref{eq:invariant}], the $w(z)$ subproblem is identical at every $\alpha$---the gradients, the anti-alignment $\cos\theta = -0.97$, and the recovered $w(z)$ curves are unchanged across the entire scan range.

The consequence is that the two operations carry almost no synergy: doing both jointly is essentially no better than doing each in turn.  The reason is geometric.  In the BAO and SN distances $\alpha$ enters only as the overall ruler normalization $D_V/(\alpha\, r_d)$---a single multiplicative constant common to every redshift---whereas $w(z)$ modifies the relative $z$-dependence of $E(z)$ that fixes the distance ratios.  Rescaling the overall amplitude and reshaping the relative ratios act on orthogonal degrees of freedom: changing the normalization does not alter which shape best fits the ratios, and reshaping the ratios does not change the normalization that best fits the amplitude.  For this distance sector the total-tension objective therefore separates additively, $T(\alpha, w) \approx f_{\rm scale}(\alpha) + h_{\rm shape}(w)$, and its minimizer is the pair $(\alpha^\ast, w^\ast)$ obtained from the two one-dimensional problems.

This clean separation is analytic only for the BAO and SN distances.  For the CMB the $\ell$-rescaling weakly mixes scale and shape through damping-tail distortions (Sec.~\ref{sec:beta}): along the Planck diagonal, lowering $\alpha$ both raises $H_0$ and lowers $\Omega_m$, so $\alpha$ does leak into the $\Omega_m$ axis that $w(z)$ nominally controls, and the cross-term $\partial^2 T/\partial\alpha\,\partial w$ is small but not strictly zero.  This residual coupling cannot help close the gap, however: it acts along the CMB peak-shift direction---degenerate with $w(z)$ and orthogonal to the BAO--SN $\Omega_m$ conflict---rather than along the gap itself.  The factorization is thus an excellent approximation rather than an exact identity, and we confirm it operationally: the optimal $\alpha^\ast = 0.992$ is already located by the $\Lambda$CDM ($w=-1$) $\alpha$-scan of Sec.~\ref{sec:results}, and the subsequent fixed-$\alpha^\ast$ $w(z)$ optimization neither shifts $\alpha^\ast$ nor narrows the irreducible $\Omega_m$ gap.  To the accuracy of this factorization the sequential optimum coincides with the joint one, and we find no value of $\alpha$ at which $w(z)$ becomes more effective.

An important caveat: this parameter-space orthogonality (scale vs.\ shape) coexists with a near-degeneracy of $\alpha$ and $w(z)$ in \emph{observable} space---$w(z)$ can reproduce almost everything $\alpha$ does to the data---and we resolve this apparent paradox in Sec.~\ref{sec:overlap}.

\section{MCMC verification}\label{sec:results}

The MCMC analysis uses MontePython \cite{Audren:2012wb,Brinckmann:2018cvx} with CLASS \cite{Blas:2011rf}, fitting six $\Lambda$CDM parameters $\{\omega_b, \omega_c, h, \tau, \ln(10^{10}A_s), n_s\}$ along with the Planck calibration nuisance parameter $A_{\rm planck}$.  For each fixed $\alpha$, we run joint and individual-dataset chains until convergence (Gelman--Rubin $R - 1 < 0.05$ for all constrained parameters) and extract best-fit $\chi^2$ values and posterior statistics.

\subsection{$\alpha$ scan and the tension trade-off}\label{sec:alpha_scan}

We scan $\alpha$ over $[0.970, 1.000]$ in steps of $\Delta\alpha = 0.001$, running chains at each value.  Table~\ref{tab:alpha_scan} shows representative points spanning the range; the SN $\chi^2$ is independent of $\alpha$ (fixed at $\chi^2_{\rm SN} = 1409.73$).  We do not extend the scan to the deeper $r_s$ reductions ($\sim$5--$8\%$) discussed in the Introduction: $T$ rises monotonically for $\alpha < \alpha^*$ (Table~\ref{tab:alpha_scan}), so larger reductions only worsen the inter-dataset tension and the relevant optimum lies well within the scanned window.

\begin{table}[t]
\caption{Representative points from the MCMC $\alpha$-scan over $[0.970, 1.000]$ (step size $\Delta\alpha = 0.001$).  $T$ is the total inter-dataset tension [Eq.~\eqref{eq:tension}], with pairwise decomposition $T_{\rm PB}$ (Planck--BAO), $T_{\rm PS}$ (Planck--SN), and $T_{\rm BS}$ (BAO--SN).  The optimal $\alpha^* = 0.992$ minimizes $T$.  All chains fit six $\Lambda$CDM parameters plus $A_{\rm planck}$.}
\label{tab:alpha_scan}
\begin{ruledtabular}
\begin{tabular}{ccccccc}
$\alpha$ & $H_0$ & $\Omega_m$ & $T_{\rm PB}$ & $T_{\rm PS}$ & $T_{\rm BS}$ & $T$ \\
\hline
0.980 & 72.0 & 0.287 & 29.8 & 20.9 & 3.4 & 40.5 \\
0.986 & 71.3 & 0.287 & 14.1 & 15.5 & 3.4 & 28.9 \\
\textbf{0.992} & \textbf{70.3} & \textbf{0.290} & \textbf{4.2} & \textbf{18.3} & \textbf{3.6} & \textbf{24.8} \\
0.995 & 69.8 & 0.292 & 1.1 & 22.6 & 4.2 & 25.6 \\
1.000 & 69.1 & 0.293 & 0.7 & 31.1 & 6.1 & 32.4 \\
\end{tabular}
\end{ruledtabular}
\end{table}

\begin{figure}[tb]
\includegraphics[width=\columnwidth]{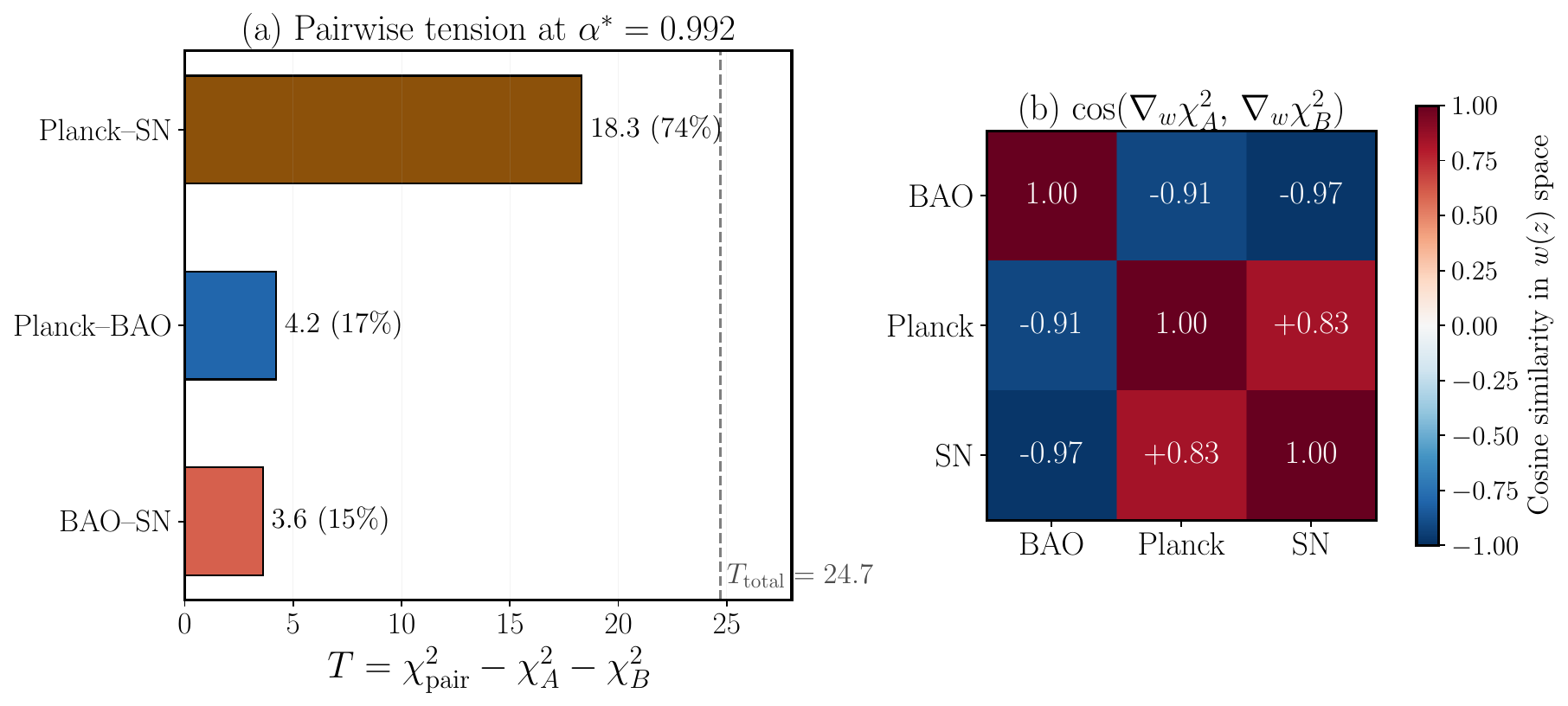}
\caption{Pairwise tension decomposition as a function of $\alpha$.  Reducing $\alpha$ trades Planck--SN tension (decreasing) for Planck--BAO tension (increasing).  BAO--SN tension is largest near $\alpha = 1$, where BAO's $H_0$ ($\approx 69$) sits farthest from SN's ($73.2$); as $\alpha$ decreases, BAO's $H_0$ rises toward SN's and $T_{\rm BS}$ shrinks, although the underlying $\Delta\Omega_m$ is exactly $\alpha$-invariant.  The total tension $T$ reaches its minimum at $\alpha^* = 0.992$.}
\label{fig:tension_decomp}
\end{figure}

The pairwise decomposition reveals a clear trade-off mechanism (Fig.~\ref{fig:tension_decomp}).  At $\alpha = 1$ ($\Lambda$CDM), the tension is dominated by $T_{\rm PS} = 31.1$ ($5.2\sigma$): Planck and SN differ in both $H_0$ ($68.6$ vs.\ $73.2$) and $\Omega_m$ ($0.300$ vs.\ $0.334$), with the large $H_0$ gap the leading contributor.  As $\alpha$ decreases, Planck's preferred $H_0$ rises (the peak shift is compensated by larger $h$), closing the $H_0$ component and reducing $T_{\rm PS}$.  But simultaneously, $\Omega_m^{\rm Planck}$ falls (through $\Omega_m h^2 \approx \text{const}$), opening a gap with BAO ($\Omega_m^{\rm BAO} = 0.297$, invariant) and increasing $T_{\rm PB}$.

The minimum at $\alpha^* = 0.992$ ($T = 24.8$, $4.0\sigma$ for $\chi^2(4)$) represents the balance point: $T_{\rm PS}$ has decreased to 18.3 ($3.9\sigma$) while $T_{\rm PB}$ has risen to 4.2 ($1.5\sigma$).  The BAO--SN $\Omega_m$ gap $\Delta\Omega_m = 0.037$ is exactly $\alpha$-invariant, as established in Sec.~\ref{sec:scale_shape}; however, the pairwise tension $T_{\rm BS}$ also includes an $H_0$ component---since BAO's $H_0$ shifts with $\alpha$ while SN's does not---so $T_{\rm BS}$ falls from 6.1 at $\alpha = 1$ to 3.4 at low $\alpha$, as BAO's $H_0$ rises toward SN's.

The joint best-fit at $\alpha^* = 0.992$ gives $H_0 = 70.3 \pm 0.3\;\mathrm{km\,s^{-1}\,Mpc^{-1}}$ and $\Omega_m = 0.290 \pm 0.004$---still $3.2\sigma$ below SH0ES.  The Planck penalty for $\alpha = 0.992$ is minimal: $\chi^2_{\rm Planck}$ rises by only 1.0 relative to $\alpha = 1$, because the full parameter freedom (particularly $n_s$, $\omega_b$, $\tau$) can partially compensate the peak shift.  This contrasts sharply with restricted analyses that fix these parameters, which vastly overestimate the Planck spectral penalty \cite{Zhou:2024dbt}.

\subsection{Best-fit trajectories in $(\Omega_m, H_0)$}\label{sec:trajectories}

\begin{figure}[tb]
\includegraphics[width=\columnwidth]{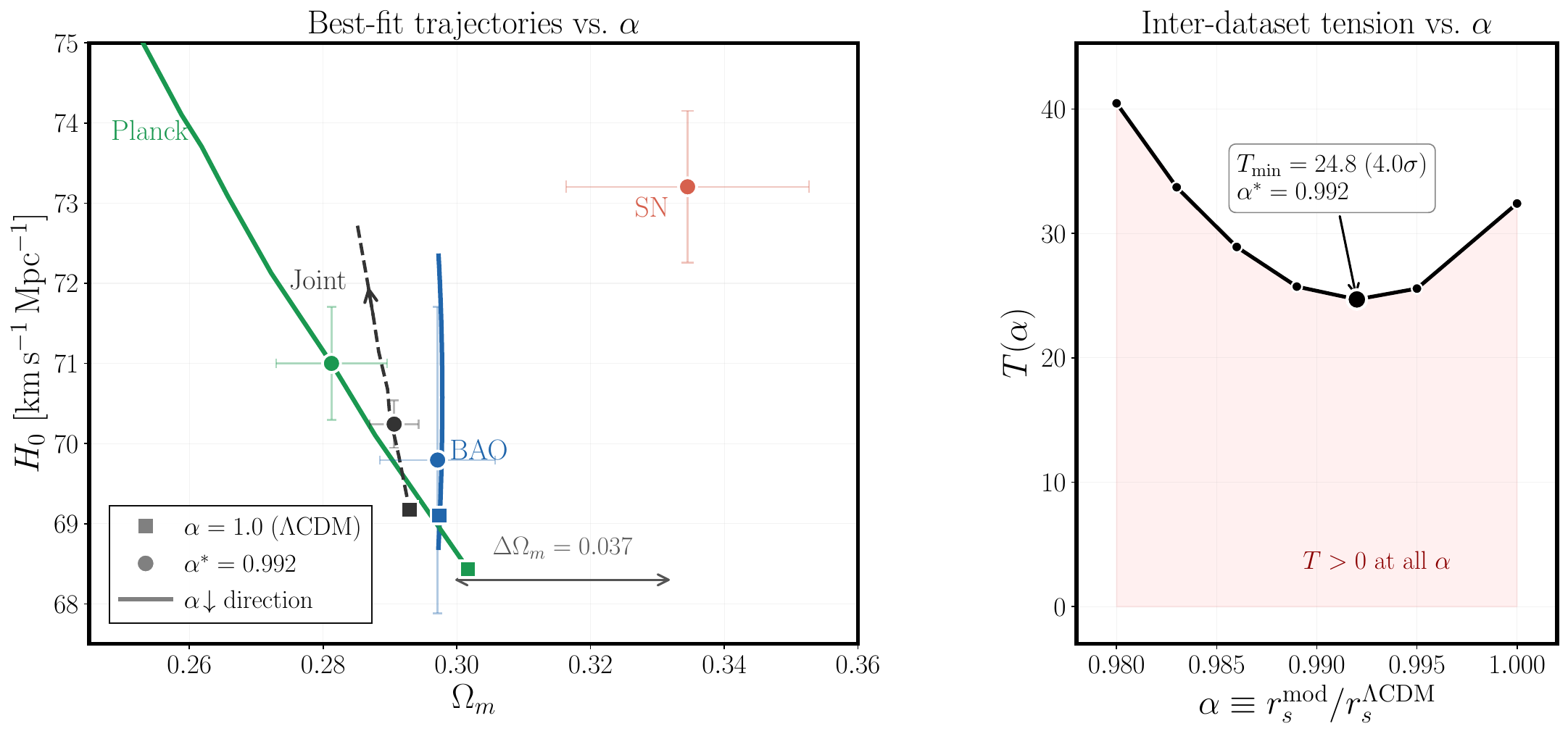}
\caption{Best-fit trajectories in $(\Omega_m, H_0)$ as $\alpha$ varies over $[0.970, 1.000]$.  BAO traces a vertical path ($\Omega_m = 0.297$ constant), SN is a fixed point, and Planck moves diagonally due to $\Omega_m h^2 \approx \text{const}$.  The BAO--SN $\Omega_m$ gap is exactly $\alpha$-invariant.}
\label{fig:trajectory}
\end{figure}

Figure~\ref{fig:trajectory} shows the per-dataset best-fit trajectories as $\alpha$ varies over $[0.970, 1.000]$, confirming the geometric predictions of Sec.~\ref{sec:scale_shape}.  The BAO best-fit traces a vertical line at constant $\Omega_m = 0.297$, with only $H_0$ responding to $\alpha$---precisely the scale--shape decoupling predicted above.  The SN best-fit is an exact fixed point, as expected from its $\alpha$-independence.  The BAO--SN gap $\Delta\Omega_m = 0.037$ is therefore constant to numerical precision ($<0.001$) at every $\alpha$, verifying Eq.~\eqref{eq:invariant}.  (The Planck $H_0 = 68.6$ at $\alpha = 1$ differs from the published value $67.36 \pm 0.54$ because our analysis uses only plik\_lite TT+TE+EE without lowE or lensing likelihoods.)

The Planck trajectory is qualitatively different: it traces a steep diagonal reflecting the $\Omega_m h^2 \approx \text{const}$ constraint.  As $\alpha$ decreases, $H_0$ rises but $\Omega_m$ falls.  At the optimal $\alpha^* = 0.992$, Planck's $\Omega_m = 0.282$ is reasonably close to BAO's 0.297 but far from SN's 0.334.  Reducing $\alpha$ further toward 0.986 would bring Planck's $H_0$ to $\sim$73, aligning with SH0ES, but at the cost of $\Omega_m^{\rm Planck} = 0.266$.  The resulting Planck--BAO discrepancy is then $\sim 3.5\sigma$---this is the \emph{full two-dimensional} $T_{\rm PB}$, driven by the $\Delta\Omega_m = 0.031$ gap together with a residual $H_0$ offset, and not by $\Omega_m$ alone (a one-dimensional $\Omega_m$ comparison at this gap would give $\approx 2.4\sigma$).  This is the fundamental trade-off: the $\Omega_m h^2$ degeneracy ensures that any $H_0$ increase comes at the price of an $\Omega_m$ decrease.

\subsection{The irreducible $\Omega_m$ landscape}\label{sec:landscape}

\begin{figure*}[t]
\includegraphics[width=\textwidth]{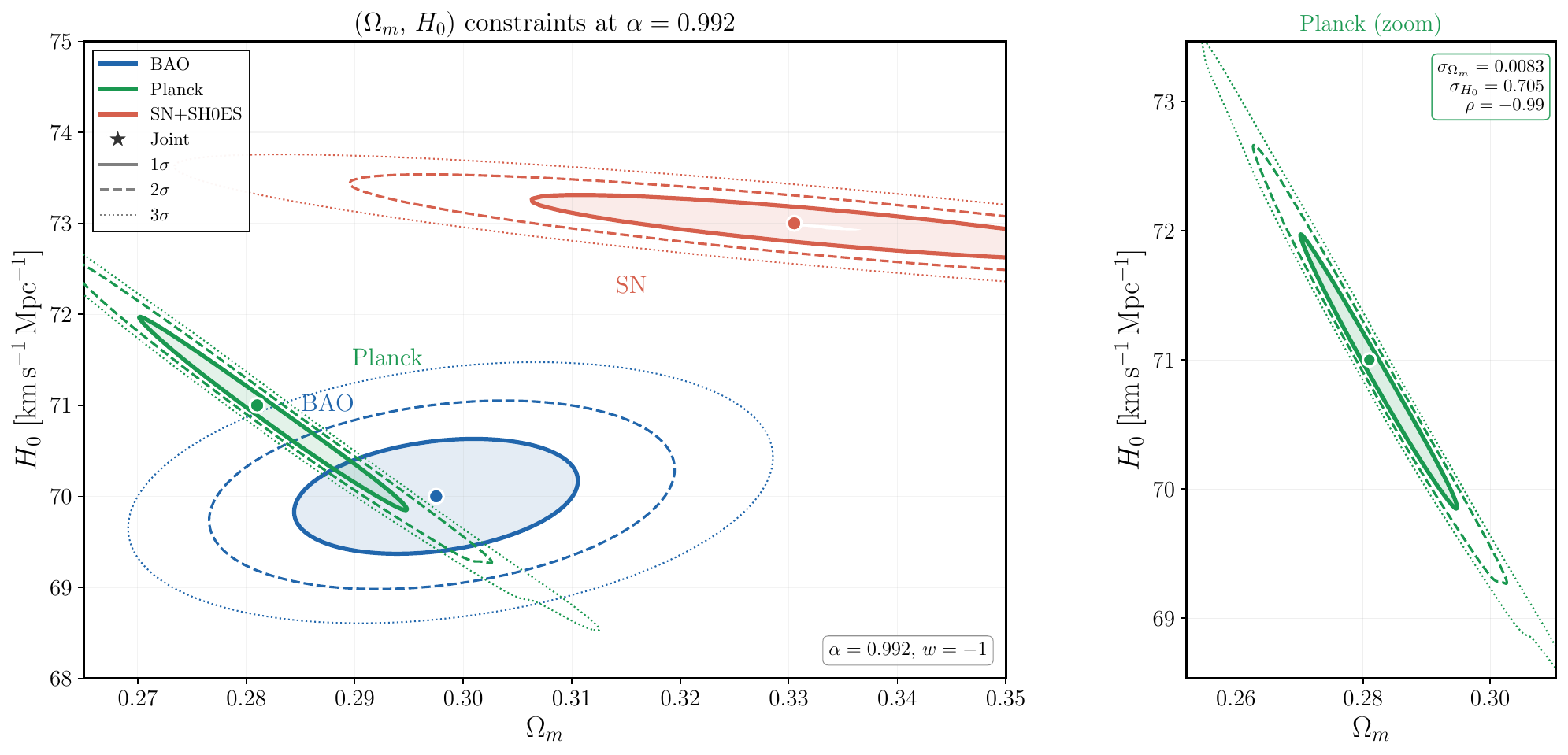}
\caption{Two-dimensional $\Delta\chi^2$ contours in the $(\Omega_m, H_0)$ plane at $\alpha^* = 0.992$.  BAO (blue), Planck (green, Fisher ellipse), and SN (red) show fully separated $1\sigma$ regions.  The joint minimum (black star) lies outside all individual preferred regions.  Right panel: zoom on the Planck ellipse.}
\label{fig:tension_surface}
\end{figure*}

At the optimal $\alpha^* = 0.992$, the per-dataset $\Omega_m$ preferences are shown in Table~\ref{tab:bestfits}.  The three datasets are separated in $\Omega_m$: Planck at $0.282$, BAO at $0.297$, and SN at $0.334$ (Fig.~\ref{fig:tension_surface}).  The pairwise gaps are:
\begin{align}
	\Delta\Omega_m(\text{Planck}, \text{BAO}) &= 0.015 \quad (T_{\rm PB} = 4.2,\; 1.5\sigma), \\
	\Delta\Omega_m(\text{BAO}, \text{SN}) &= 0.037 \quad (T_{\rm BS} = 3.6,\; 1.4\sigma), \\
	\Delta\Omega_m(\text{Planck}, \text{SN}) &= 0.052 \quad (T_{\rm PS} = 18.3,\; 3.9\sigma).
\end{align}
The dominant tension is between Planck and SN; at $\alpha^* = 0.992$ the $H_0$ gap has shrunk to $2.2\;\mathrm{km\,s^{-1}\,Mpc^{-1}}$ (Planck $71.0$ vs.\ SN $73.2$), so the residual $T_{\rm PS} = 18.3$ is now driven by their $0.052$ gap in $\Omega_m$ combined with Planck's extremely tight constraints (613 bins).  The joint best-fit $\Omega_m = 0.290$ lies between Planck and BAO but far from SN---an unsatisfactory compromise that no individual dataset prefers.

The joint $H_0 = 70.3$ may at first appear anomalously low, since it falls below all three individual values (Planck $71.0$, BAO $70.6$, SN $73.2$), whereas a naive inverse-variance average of the three would give $\approx 71.6$.  This is not an inconsistency but a direct consequence of Planck's tight $\Omega_m$--$H_0$ degeneracy.  Planck dominates the joint $H_0$ constraint (its $\pm 0.7$ posterior is far tighter than BAO's $\pm 1.9$ or SN's $\pm 1.0$), so the joint $H_0$ is set essentially by where Planck's degeneracy line crosses the compromise $\Omega_m$, not by an average of the three central values.  Because the joint fit is pulled to $\Omega_m = 0.290$, slightly \emph{above} Planck's own preferred $0.282$, moving up Planck's degeneracy ($\Omega_m h^2 \approx \text{const}$, so $\Omega_m\!\uparrow \Rightarrow H_0\!\downarrow$) by this $\Delta\Omega_m = 0.008$ lowers $H_0$ below Planck's $71.0$.  The joint value therefore lies below every individual best-fit by construction.

\begin{table}[b]
\caption{Individual dataset best-fits at $\alpha^* = 0.992$ from MCMC.  $H_0$ in $\mathrm{km\,s^{-1}\,Mpc^{-1}}$.  Uncertainties are posterior $1\sigma$.  The $\Omega_m$ ordering Planck $<$ BAO $<$ SN is the source of the residual tension.  The SN $H_0$ reflects the SH0ES Cepheid calibration of $M_B$; after marginalizing $M_B$ with this Gaussian prior, $\Omega_m$ is determined by relative distances while $H_0$ is anchored by the local distance ladder.}
\label{tab:bestfits}
\begin{ruledtabular}
\begin{tabular}{lccc}
Dataset & $\Omega_m$ & $H_0$ & $\chi^2_{\min}$ \\
\hline
Planck & $0.282 \pm 0.009$ & $71.0 \pm 0.7$ & 543.4 \\
BAO & $0.297 \pm 0.009$ & $70.6 \pm 1.9$ & 10.3 \\
SN & $0.334 \pm 0.018$ & $73.2 \pm 1.0$ & 1409.7 \\
\hline
Joint & $0.290 \pm 0.004$ & $70.3 \pm 0.3$ & 1988.2 \\
\end{tabular}
\end{ruledtabular}
\end{table}

\section{Why $w(z)$ cannot close the gap}\label{sec:wz}

\subsection{$w(z)$ recovery curves}\label{sec:recovery}

\begin{figure}[t]
\includegraphics[width=\columnwidth]{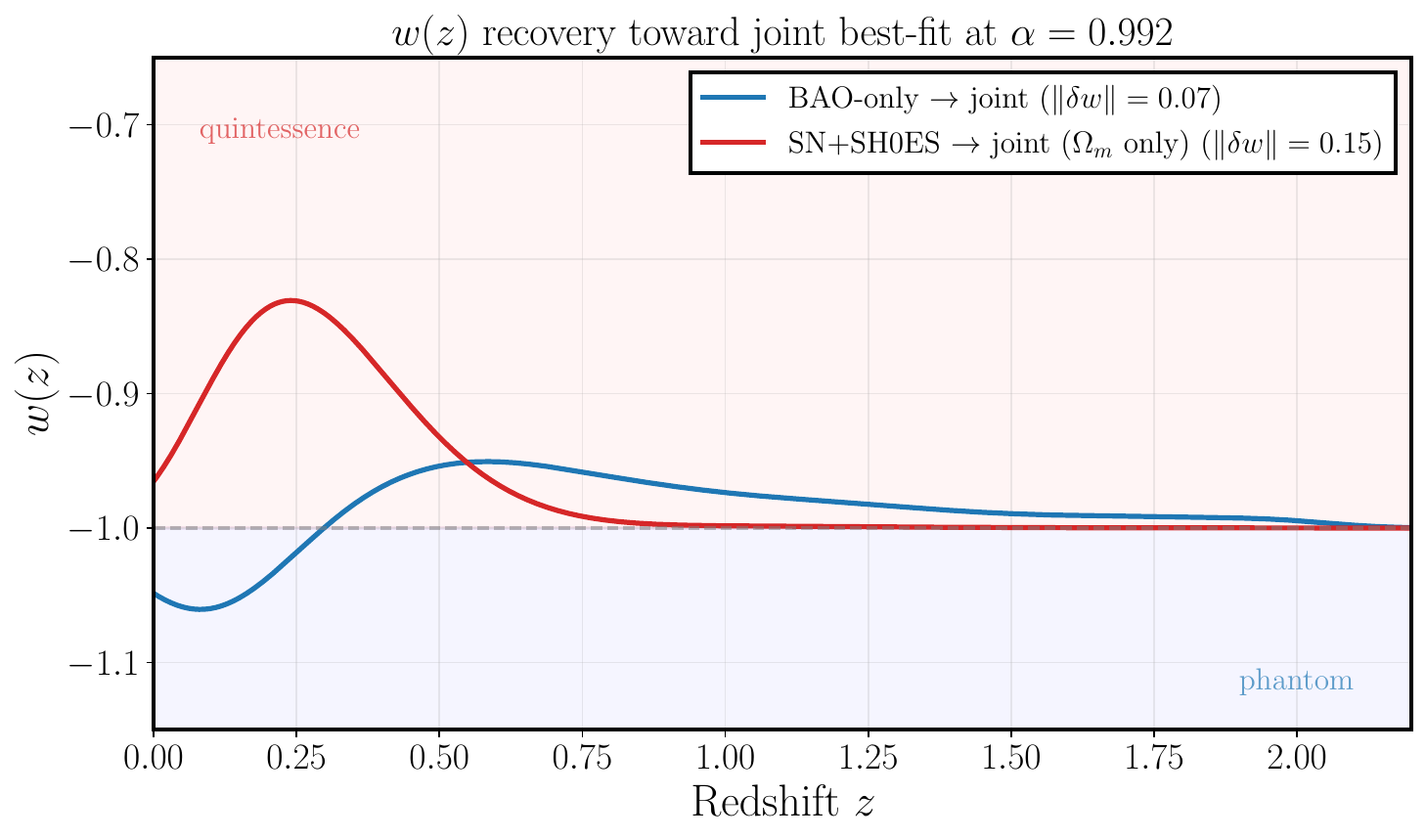}
\caption{Dark energy equation of state $w(z)$ required by each dataset to reach the joint best-fit from its own optimum, at $\alpha^* = 0.992$.  BAO (blue) needs phantom dark energy ($w < -1$) at low redshift, while SN (red) needs quintessence ($w > -1$).  The curves lie on opposite sides of $\Lambda$CDM ($w = -1$, dashed), confirming the gradient anti-alignment.}
\label{fig:wz_recovery}
\end{figure}

To visualize the anti-alignment, we invert the FPA response relation [Eq.~\eqref{eq:response}] to solve for the $w(z)$ coefficients that would shift each dataset's best-fit parameters to the joint optimum (Fig.~\ref{fig:wz_recovery}).  At $\alpha^* = 0.992$:

Both curves are constructed to isolate the \emph{shape} ($\Omega_m$) component of the displacement, so that the comparison is symmetric.  For BAO the accompanying scale shift is negligible ($\Delta H_0 = -0.3\;\mathrm{km\,s^{-1}\,Mpc^{-1}}$) and leaves the curve essentially unchanged; for SN we explicitly hold $H_0$ fixed and move $\Omega_m$ alone.  The infeasibility of the much larger SN \emph{scale} ($H_0$) displacement is a separate matter, treated in the following paragraph.

The BAO recovery ($\Delta\Omega_m = -0.007$, $\Delta H_0 = -0.3\;\mathrm{km\,s^{-1}\,Mpc^{-1}}$) requires only a mild phantom deformation, $w \approx -1.05$ at $z \sim 0$, which reshapes the distance--redshift relation to accommodate the shift from BAO's preferred $\Omega_m = 0.297$ to the joint value $0.290$.  The SN recovery ($\Delta\Omega_m = -0.044$) pulls in the opposite direction: with $H_0$ held at SN's value ($73.2\;\mathrm{km\,s^{-1}\,Mpc^{-1}}$) the recovery reduces to $\Delta\Omega_m$ alone and yields a moderate quintessence deformation, $w \approx -0.83$ at $z \sim 0.25$, smoothly shifting SN's preferred $\Omega_m$ from $0.334$ to $0.290$.

The two curves lie on opposite sides of $w = -1$ in the most constraining redshift range ($z < 0.5$), crossing near $z \approx 0.5$.  A single smooth $w(z)$ cannot be simultaneously phantom and quintessence in the same redshift range.  This is the visual manifestation of the gradient anti-alignment $\cos\theta = -0.97$ reported in Eq.~\eqref{eq:antialign}.

The SH0ES Cepheid calibration pins $H_0 = 73.2\;\mathrm{km\,s^{-1}\,Mpc^{-1}}$ through the low-redshift anchor, so the full SN recovery must simultaneously accommodate $\Delta H_0 = -2.9\;\mathrm{km\,s^{-1}\,Mpc^{-1}}$ (to reach the joint $H_0 = 70.3$) in addition to $\Delta\Omega_m = -0.044$.  The resulting $w(z)$ exhibits violent oscillations ($w$ ranging from $+2.3$ to $-4.5$), a manifestly non-physical solution.  The failure is structural, and the way it sets in is instructive.  Because $E(z) \to 1$ as $z \to 0$ for \emph{any} $w(z)$, the low-redshift Hubble-flow slope that pins $H_0$ at fixed $M_B$ is insensitive to dark-energy reshaping; $w(z)$ can only redistribute distances at higher redshift.  The SN-inferred $H_0$ therefore responds only weakly to $w(z)$, and the coefficients needed to shift it scale inversely with this small sensitivity: a modest $\Delta H_0$ is absorbed by a mild, smooth deformation, but a larger one demands coefficients so big that the deformation leaves the physical regime (where $E(z) > 0$ everywhere).  Pushing all the way to $\Delta H_0 = -2.9$ then drives the recovery into the near-degenerate, poorly constrained directions of the response operator, and the optimizer fills them with high-frequency oscillations---an artifact of inverting a near-singular map, not a physical solution.

The root cause is geometric.  A late-time $w(z)$ reshapes the distance--redshift relation through the integrand of $d_L(z) \propto (c/H_0)\int_0^z dz'/E(z')$ but cannot rescale the overall amplitude $c/H_0$, which is exactly what an $H_0$ shift at fixed $M_B$ requires.  Late-time $w(z)$ modifications therefore cannot substitute for a recalibration of the local distance ladder; expressed through $H_0$ directly, the Hubble tension is fundamentally an absolute-scale problem that lies outside the reach of shape deformations.

\subsection{Nonlinear verification: CPL parametrization}\label{sec:cpl}

To confirm that the anti-alignment is not a linearization artifact, we perform direct nonlinear optimization within the CPL parametrization $w(a) = w_0 + w_a(1-a)$ at fixed $\alpha = 0.992$, fitting $\{\omega_c, h, w_0, w_a\}$ simultaneously (Table~\ref{tab:cpl}).  Holding $\{\omega_b, \tau, n_s, A_s\}$ fixed here is a conservative choice for this directional cross-check: freeing them would only give CPL additional room to lower $\chi^2$, strengthening rather than weakening the conclusion that no $w(z)$ reconciles the probes.

\begin{table}[t]
\caption{CPL $w_0 w_a$CDM fits at fixed $\alpha = 0.992$, with $\{\omega_b, \tau, n_s, A_s\}$ fixed.  $\Delta\chi^2$ is relative to $\Lambda$CDM ($w_0 = -1$, $w_a = 0$) at the same $\alpha$ and with the same fixed parameters.}
\label{tab:cpl}
\begin{ruledtabular}
\begin{tabular}{lcccc}
Dataset & $w_0$ & $w_a$ & $w(z{=}0.5)$ & $\Delta\chi^2$ \\
\hline
BAO & $-0.18$ & $-2.71$ & $-1.08$ & $-4.7$ \\
SN & $-0.89$ & $0.00$ & $-0.89$ & $-0.5$ \\
Planck & $-1.00$ & $0.00$ & $-1.00$ & $-0.1$ \\
BAO+SN & $-1.09$ & $0.00$ & $-1.09$ & $-21.2$ \\
\end{tabular}
\end{ruledtabular}
\end{table}

Neither SN ($\Delta\chi^2 = -0.5$) nor Planck ($\Delta\chi^2 = -0.1$) individually benefits from $w \neq -1$; both are well-fit by $\Lambda$CDM.  The BAO-only fit finds $w_0 = -0.18$ (quintessence at $z = 0$, evolving to phantom by $z \gtrsim 0.4$), but this extreme CPL trajectory is driven by only 13 data points constraining 4 free parameters---the constraint is very weak and the $w_a = -2.71$ carries large uncertainty.  Importantly, this BAO ``absolute preference'' for quintessence at $z = 0$ does not contradict the recovery analysis of Sec.~\ref{sec:recovery}, which asks a different question: what $w(z)$ deformation is needed to move BAO from its own $\Lambda$CDM optimum \emph{toward the joint best-fit}.  That displacement requires phantom $w$, because BAO must decrease its $\Omega_m$ from 0.297 to 0.290.

We do not report a full joint (BAO+SN+Planck) CPL fit because, with all three datasets combined, $w_0w_a$ primarily absorbs inter-dataset tension rather than reflecting a physical dark energy preference.  The critical entry is BAO+SN: the combined fit requires phantom dark energy ($w_0 = -1.09$, $\Delta\chi^2 = -21.2$), which is the nonlinear manifestation of the gradient anti-alignment.  The phantom $w$ partially closes the $\Omega_m$ gap by shifting both datasets toward an intermediate value, but it pushes the equation of state away from what either dataset individually prefers.

\section{Effective freedom of the deformation space}\label{sec:freedom}

The 22-parameter space $\{\alpha, \beta_{\rm damp}, w_1, \ldots, w_{20}\}$ might suggest large freedom to resolve conflicts.  This section shows that the effective cosmological freedom is much smaller.

\subsection{Response overlap and the apparent contradiction}\label{sec:overlap}

99.8\% of the $\alpha$ response lies within the $w(z)$ response subspace: $\|P_w R_\alpha\|^2 / \|R_\alpha\|^2 = 0.998$.  In plain terms, $w(z)$ can reproduce almost everything $\alpha$ does to the data.  At first sight, this seems to contradict the factorization theorem---if $\alpha$ produces observable signatures almost entirely replicable by $w(z)$, how can they act on ``orthogonal directions''?

The resolution is that the factorization operates in the two-dimensional $(\Omega_m, H_0)$ parameter space, while the 99.8\% overlap is measured in the 2206-dimensional observable space.  Both $\alpha$ and $w(z)$ modify the same distance observables (hence the high overlap), but they \emph{control different parameters}: $\alpha$ adjusts the ruler normalization ($H_0$) while leaving the ruler-independent shape ($\Omega_m$) invariant.  The 0.2\% non-overlapping component corresponds to the CMB-specific spectral signature of peak shifting that $w(z)$ cannot replicate.

The high overlap actually \emph{strengthens} the obstruction: the nominal 22 parameters collapse to only $\sim$3 independent observable directions (at 90\% variance threshold in the normalized Gram matrix; $\sim$5 at 95\%), so the apparent freedom is largely illusory.  The 0.2\% of $\alpha$'s response that $w(z)$ cannot reproduce is the CMB-specific peak shift---the very lever behind the Planck trade-off---but it does nothing to close the BAO--SN $\Omega_m$ gap.

This connects to Lee's finding \cite{Lee:2026geo} that extending $\Lambda$CDM to wCDM reduces the leading Planck Fisher eigenvalue by a factor of $37.5$: the curvature compression in Fisher space is the dual of our response-vector overlap in observable space.  Both analyses, from completely different starting points, reach the same conclusion: extending $w(z)$ does not increase effective cosmological freedom.

\subsection{Coverage of the $\Omega_m$ direction}\label{sec:coverage}

Is the obstruction caused by insufficient model freedom---an inability of $\{\alpha, \beta_{\rm damp}, w(z)\}$ to reach the $\Omega_m$ observable direction?  No.  The projection of the $\Omega_m$ response vector onto span$(\alpha, \beta, w)$ is $93\%$---the deformation modes can reach the $\Omega_m$ direction with high fidelity.  The obstruction is not ``we cannot get there'' but ``the datasets disagree about \emph{where} to go.''  Of this 7\% unreachable component, two-thirds is Planck-internal, arising from matter--radiation equality spectral features that smooth distance modifications cannot mimic.

\subsection{Damping deformation $\beta_{\rm damp}$}\label{sec:beta}

The damping parameter $\beta_{\rm damp}$ is an independent observable direction: it is 100\% Planck-internal, nearly orthogonal to $\alpha$ ($\cos\theta = 0.10$ in observable space), and has 27\% content outside span$(R_w)$.  However, it operates purely within the CMB spectral space and cannot affect the BAO--SN $\Omega_m$ conflict.  MCMC verification at $\alpha^* = 0.992$ shows that $\beta_{\rm damp}$ monotonically worsens the fit (Table~\ref{tab:beta}): with full parameter freedom, the lensed Planck spectrum is already well-fit ($\chi^2/{\rm dof} = 0.90$), leaving no room for damping modifications.

\begin{table}[t]
\caption{$\beta_{\rm damp}$ scan at fixed $\alpha^* = 0.992$.  $\beta_{\rm damp} = 0$ is optimal; any nonzero value worsens the joint fit.}
\label{tab:beta}
\begin{ruledtabular}
\begin{tabular}{ccc}
$\beta_{\rm damp}$ ($\times 10^{-9}$) & $\chi^2_{\rm joint}$ & $\Delta\chi^2$ \\
\hline
0 & 1988.2 & 0.0 \\
3 & 1989.3 & +1.1 \\
9 & 1991.9 & +3.7 \\
20 & 2007.5 & +19.3 \\
\end{tabular}
\end{ruledtabular}
\end{table}

\section{Robustness}\label{sec:robust}

\subsection{SN dataset independence}\label{sec:dessn}

\begin{figure}[t]
\includegraphics[width=\columnwidth]{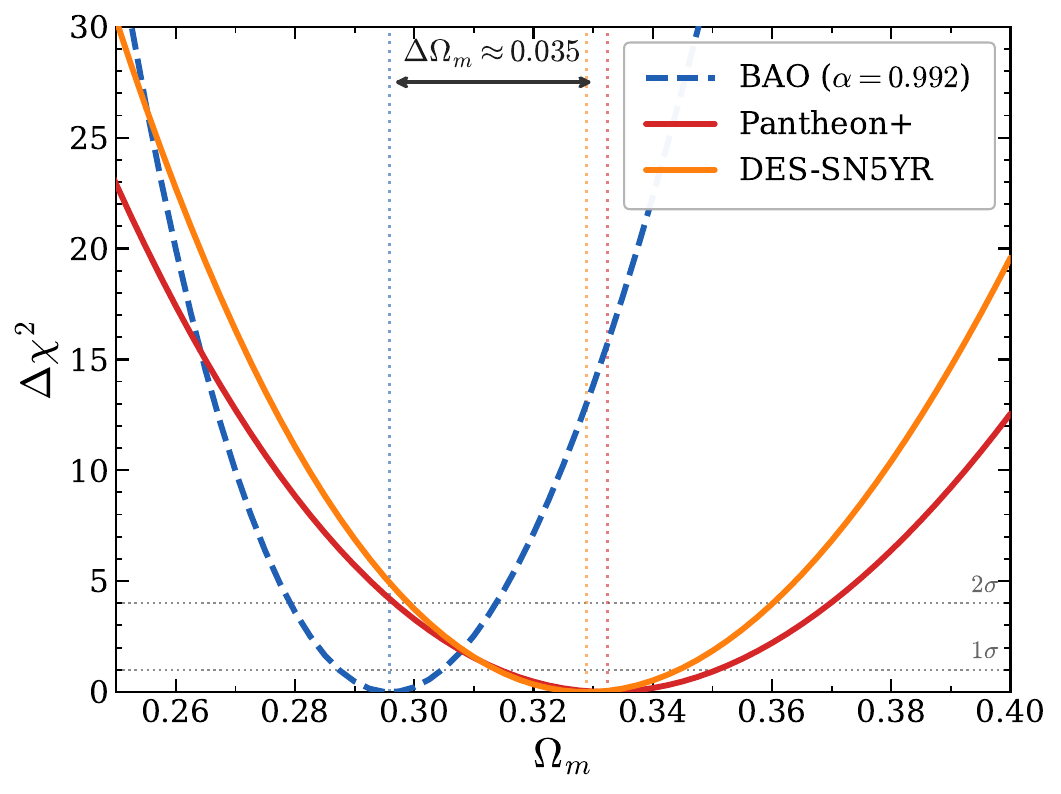}
\caption{$\Omega_m$ profile $\chi^2$ for BAO (blue), Pantheon+ (red), and DES-SN5YR (orange), with $H_0$ profiled out.  Both SN datasets prefer $\Omega_m \approx 0.330$, separated from BAO's $\Omega_m = 0.297$ by $\Delta\Omega_m \approx 0.035$.  The obstruction is SN-dataset independent.}
\label{fig:dessn}
\end{figure}

Replacing Pantheon+ with DES-SN5YR (1820 SNe) yields $\Omega_m^{\rm SN} = 0.330$, giving $\Delta\Omega_m(\text{BAO},\text{SN}) = 0.035$---in close agreement with the Pantheon+ value (0.037) (Fig.~\ref{fig:dessn}).  The BAO--SN $\Omega_m$ discrepancy is confirmed independently by \'O~Colg\'ain and Sheikh-Jabbari \cite{OColgain:2024xqj}, who find $\Omega_m \approx 0.33$ for both DES and Pantheon+ through the $w_0w_a \to \Lambda$CDM mapping.

\subsection{Gradient metric dependence}\label{sec:metric_robust}

The BAO--SN gradient anti-alignment is measured in a 20-dimensional $w(z)$ space, and its quantitative value depends on the choice of inner product.  Under the Euclidean norm, $\cos\theta = -0.97$; under the Fisher-weighted ($F^{-1}$) norm, the anti-alignment weakens to $\cos\theta = -0.34$.  The reduction reflects the Fisher weighting's emphasis on parameter directions to which the data are most sensitive, effectively projecting out poorly constrained modes.

The sign---negative cosine, indicating opposing gradient directions---is the load-bearing feature, not the magnitude.  A near-antiparallel $\cos\theta = -0.97$ means the cone of directions that improve both datasets is vanishingly narrow, and---crucially---closing the $\Omega_m$ gap requires deformations of \emph{opposite sign} for the two probes (Sec.~\ref{sec:recovery}): BAO needs phantom ($w<-1$) while SN needs quintessence ($w>-1$) at $z<0.5$.  Fully reconciling them would require a positive projection, which neither metric produces.  The weaker $F^{-1}$-weighted value ($\cos\theta = -0.34$) indicates that in parameter-relevant directions the opposing pulls are less extreme---but they are still opposing, and the sign conflict in the required $w(z)$ deformations persists.

\subsection{Sensitivity to fixed parameters}\label{sec:fixed_params}

The MCMC analysis fits all six $\Lambda$CDM parameters, but the FPA gradient analysis fixes $\{\omega_b, \tau, n_s, A_s\}$.  This is justified because:
\begin{itemize}
	\item $\omega_b$ primarily affects distances through $r_s$ and baryon loading.  The $r_s$ component is absorbed by $\alpha$.  Baryon loading modifies relative CMB peak heights, but the 613-bin Planck spectrum constrains $\omega_b$ to $\pm 0.3\%$, producing negligible shifts in $\Omega_m$ ($\delta\Omega_m \lesssim 0.001$).
	\item $\tau$ and $A_s$ are degenerate and affect only the CMB amplitude envelope, not the distance observables that drive inter-dataset tension.
	\item $n_s$ affects the CMB spectral tilt and has a weak correlation with $\Omega_m$ through the damping tail.  Planck constrains $n_s$ to $\pm 0.004$; the resulting $\Omega_m$ shift ($\delta\Omega_m \lesssim 0.003$) cannot close the 0.037 gap.
\end{itemize}
The MCMC validation, which frees all these parameters, confirms that the qualitative conclusions---$\alpha$-invariance of $\Delta\Omega_m$, gradient anti-alignment, and the tension trade-off---are unchanged.

\subsection{Scope of the obstruction}\label{sec:scope}

Our result applies to early-time sound horizon reduction (parametrized by $\alpha$) combined with late-time smooth $w(z)$ modifications of the expansion history.  It does not exclude three broad classes of physics that lie outside this space.  First, perturbation-level effects---clustering dark energy ($c_s^2 \neq 1$), modified gravity ($\mu$, $\Sigma$), or varying $G_{\rm eff}$---can differentially affect how each dataset constrains $\Omega_m$ through growth-rate and lensing channels, operating outside the $H(z)$-only deformation space.  Second, non-smooth features such as phase transitions, sharp $w(z)$ features, or void models can violate the smoothness assumed in our $w(z)$ basis.  Third, calibration systematics in distance-ladder anchoring or SN standardization could shift $\Omega_m^{\rm SN}$ directly.  The key distinction is that early-time $r_s$ reduction and late-time smooth $w(z)$ operate through $H(z)$ alone, whereas these modifications can change the \emph{relative} $\Omega_m$ constraints between probes.

\section{Discussion}\label{sec:discussion}

\subsection{Converging evidence}

The six contemporary analyses cited in the Introduction each map onto specific aspects of our framework.  Lee \textit{et al.}~\cite{Lee:2025yah} ($P(k)$ optimization giving $\Omega_m = 0.247$ at $H_0 = 73$) correspond to the extreme-$\alpha$ limit of the Planck diagonal trajectory.  Mirpoorian \textit{et al.}~\cite{Mirpoorian:2024fka} ($\Omega_m = 0.286$--$0.298$ with modified recombination, $k_D/r_*$ barely changed) operate within our $\{\alpha, \beta_{\rm damp}\}$ subspace.  Bansal and Huterer \cite{Bansal:2026late} (smooth $H(z)$ yields $\Delta\chi^2 \sim 0$; only SN $M_B$ transitions help) confirm the $\Omega_m$ floor via MCMC.  Pedrotti \textit{et al.}~\cite{Pedrotti:2024kpn} ($\delta\omega_c/\omega_c \approx 2.38\,\delta h/h$) identify the structural constraint behind our Planck diagonal trajectory.  Lee \cite{Lee:2026geo} (Fisher eigenvalue reduction by a factor of $37.5$ in wCDM) provides the information-geometric dual of our 99.8\% response overlap.  \'O~Colg\'ain and Sheikh-Jabbari \cite{OColgain:2024xqj} ($\Omega_m(z)$ inconsistency in DESI $w_0w_a$CDM) independently corroborate the SN--BAO $\Omega_m$ conflict.

Together with our previous gradient analysis \cite{Zhou:2024dbt}, these complementary analyses---using distinct methods and datasets---are consistent with the same conclusion: the incompatibility between distance probes---in both $\Omega_m$ and $H_0$---is the fundamental obstruction, and early-time $r_s$ reduction combined with late-time smooth $w(z)$ cannot resolve it.

\subsection{Implications}

Our results add to the growing evidence that smooth background-level modifications face a hard floor: pure late-time $w(z)$ is insufficient \cite{Zhou:2024dbt}, early-time $r_s$ reduction alone does not achieve concordance \cite{Lee:2025yah,Mirpoorian:2024fka}, and even combining both freedoms leaves residual $\Omega_m$ and $H_0$ conflicts (this work).  Within the class of smooth early-plus-late modifications explored here, the available parameter freedom does not align with the inter-dataset discrepancy.

The $\Delta\Omega_m = 0.037$ invariant and the $H_0 = 70.3\;\mathrm{km\,s^{-1}\,Mpc^{-1}}$ ceiling serve as diagnostic benchmarks: any proposed resolution must either (i) close the $\Omega_m$ gap through physics that differentially reshapes how BAO and SN constrain matter density, and bridge the $H_0$ deficit through mechanisms beyond smooth $w(z)$, or (ii) explain why these gaps reflect systematic effects rather than physical discrepancies.  Concrete directions include perturbation-level dark energy ($c_s^2 \neq 1$, which modifies BAO reconstruction through growth effects), modified gravity (which changes the lensing--$\Omega_m$ degeneracy), or a reassessment of SN calibration systematics.

An important extension concerns full early-dark-energy models, which differ from our phenomenological $\alpha$ rescaling in a key respect: the structural constraint $\delta\omega_c/\omega_c \approx 2.38\,\delta h/h$ \cite{Pedrotti:2024kpn} means that raising $h$ requires raising $\omega_c$, partially compensating the $\Omega_m$ decrease observed in our MCMC $\alpha$-scan.  However, the higher $\omega_c$ worsens the $S_8$ tension \cite{Hill:2020osr,Vagnozzi:2019ezj}, and the BAO--SN $\Delta\Omega_m$ gap---determined by relative distance ratios independent of the physical matter density $\omega_m \equiv \Omega_m h^2$---remains unaffected.  A quantitative analysis of how EDE's $\omega_m$ compensation interacts with the geometric obstruction identified here is left to future work.

\section{Conclusions}\label{sec:conclusions}

We have identified a geometric obstruction to resolving the Hubble tension through early-time sound horizon reduction combined with late-time smooth dark energy modifications.  The obstruction has three interlocking components.

The first is a scale--shape decoupling: sound horizon rescaling ($\alpha$) operates on the absolute distance scale while $\Omega_m$ is determined by relative distance ratios.  Within $\Lambda$CDM, the BAO--SN gap $\Delta\Omega_m = 0.037$ is an exact $\alpha$-invariant, confirmed by MCMC across the full range $\alpha \in [0.970, 1.000]$.

The second is the Planck trade-off.  MCMC scans over $\alpha \in [0.970, 1.000]$ reveal that reducing $\alpha$ trades Planck--SN tension for Planck--BAO tension.  The optimal $\alpha^* = 0.992$ minimizes total tension ($T = 24.8$, $4.0\sigma$) by balancing Planck--SN ($3.9\sigma$) against Planck--BAO ($1.5\sigma$), but achieves only $H_0 = 70.3 \pm 0.3\;\text{km\,s}^{-1}\,\text{Mpc}^{-1}$---still $3.2\sigma$ below SH0ES.

The third is $w(z)$ anti-alignment.  Late-time $w(z)$ deformations cannot close the $\Omega_m$ gap because BAO and SN require opposite modifications ($\cos\theta = -0.97$): moving BAO toward the joint best-fit requires phantom ($w < -1$) while moving SN toward the joint requires quintessence ($w > -1$) at $z < 0.5$.  The effective cosmological freedom of $\{\alpha, \beta_{\rm damp}, w(z)\}$ is far smaller than the nominal 22 parameters suggest: 99.8\% of the $\alpha$ response lies within the $w(z)$ response subspace, and the independent directions do not align with the inter-dataset conflict.

These conflicts are not limitations of model flexibility but properties of the data.  The deformation modes cover 93\% of the $\Omega_m$ response direction; nonetheless, BAO and SN constrain $\Omega_m$ through geometrically independent channels and disagree, while the $H_0$ deficit persists because late-time $w(z)$ modifies only relative distances, not the absolute scale.  Resolving the Hubble tension likely requires physics beyond early-time $r_s$ reduction and late-time smooth $w(z)$ engineering---either perturbation-level effects, non-smooth features, or a reassessment of calibration systematics.

A final word on scope and outlook.  Our $\alpha$ is a deliberately model-agnostic proxy for early-time sound-horizon reduction: it captures the geometric effect common to early dark energy, extra relativistic species, and modified recombination \cite{Poulin:2023lkg,Kamionkowski:2022pkx,Mirpoorian:2024fka} without committing to the microphysics of any one of them.  This is the source of both the generality of the obstruction and the route to evading it---because the obstruction follows from how a pure scale rescaling acts on distance observables, a concrete model can escape it only where it departs from that reduction.  Two such departures stand out.  Refined early-time models that simultaneously shift the matter density---through EDE's structural correlation $\delta\omega_c/\omega_c \approx 2.38\,\delta h/h$ \cite{Pedrotti:2024kpn} (Sec.~\ref{sec:discussion})---move the Planck trajectory in the $(\Omega_m, H_0)$ plane and may relax the $H_0$ ceiling, albeit at the cost of $S_8$ \cite{Hill:2020osr,Vagnozzi:2021gjh} and without affecting the relative-ratio BAO--SN gap.  More fundamentally, scenarios that \emph{couple} the early and late sectors---interacting dark energy or dark-sector energy--momentum exchange \cite{Silva:2025hxw}---lie outside our $H(z)$-only deformation space and can in principle reshape how each probe constrains $\Omega_m$, acting precisely on the relative discrepancy that smooth $w(z)$ leaves untouched.  Whether such early--late synergy \cite{Poulin:2024ken,Vagnozzi:2023nrq} can convert the geometric obstruction quantified here into a genuine resolution---rather than merely relocating the tension---is the natural target for future work.

\begin{acknowledgments}
Supported in part by Natural Science Basic Research Plan in Shaanxi Province of China (Grant No.\ 2025JC-YBQN-497) and the High-level Talents Program of Xi'an International University (Grant No.\ XAIU202518).
\end{acknowledgments}

\section*{Data Availability}
The code and data that support the findings of this article are openly available \cite{GOdata}.

\bibliography{references}

\end{document}